\documentclass[11pt]{article}
\usepackage[no-math]{fontspec}
\usepackage[a4paper,margin=2.4cm]{geometry}
\usepackage{amsmath,amssymb,amsthm,bm}
\usepackage{booktabs,array,longtable,tabularx}
\usepackage{algorithm,algpseudocode}
\usepackage{tikz,pgfplots}
\usetikzlibrary{arrows.meta,positioning}
\usepgfplotslibrary{groupplots}
\pgfplotsset{compat=1.18,every axis/.append style={tick label style={font=\small},label style={font=\small},title style={font=\small}}}
\usepackage{flafter,placeins}
\PassOptionsToPackage{hyphens}{url}
\usepackage[hidelinks]{hyperref}
\usepackage{url}
\newtheorem{proposition}{Proposition}
\newtheorem{definition}{Definition}

\newcommand{\Top}{\operatorname{Top}}
\newcommand{\ind}{\mathbf{1}}
\newcommand{\best}[1]{\textbf{#1}}
\hypersetup{pdfauthor={Long Wang},pdftitle={Context-Triggered Associative Retrieval through Lexical Conditioning}}
\title{Context-Triggered Associative Retrieval\\through Lexical Conditioning}
\author{Long Wang\\
  {\small\href{mailto:alueas@126.com}{\texttt{alueas@126.com}}}\\
  {\small\href{https://orcid.org/0009-0006-8371-1562}{ORCID: 0009-0006-8371-1562}}}
\date{}
\begin{document}
\maketitle

\begin{abstract}
A past experience can suggest a useful intervention even when it shares little wording with the current situation. A global embedding ranking may miss such an experience if candidates with similar wording occupy the highest ranks. We use the current situation to trigger associative retrieval and condition the ranking on lexical similarity. Hard conditioning gives priority to candidates with low lexical scores and ranks them by embedding similarity within that group. Soft conditioning subtracts a baseline estimated from lexical scores. Subtraction can reverse a useful ordering if the higher-ranked candidate receives a larger correction and the difference exceeds its original lead. With a frozen encoder and ten selections per context, hard conditioning retrieved 21 of 24 relevant query--history pairs across four constructed contexts covering two intervention patterns. Dense ranking and reciprocal rank fusion retrieved none, and inverse BM25 retrieved four. Replacing two unrelated candidates per context with histories that used a different intervention reduced the count to 19. The two new misses both concerned recovering evidence of completed work. In a shared library, hard conditioning missed all six relevant pairs with high lexical scores, motivating a complementary channel for histories with similar wording.

\end{abstract}

\section{Introduction}
An agent confronting a new situation may benefit from an earlier experience before it knows what to ask its memory. When six households arrive at a food pantry within the same narrow window, an experience of staggering theater coat collection can suggest a useful intervention: distribute arrivals while retaining a single attendant. A history about inaccessible donation paperwork may resemble the pantry situation more closely while addressing a different cause. Retrieving the useful experience therefore requires access to a transferable relation that need not dominate overall textual similarity.

Retrieval-augmented generation makes external information available to language models \cite{lewis2020rag}, and dense encoders allow responses or histories to be ranked by similarity \cite{karpukhin2020dpr,reimers2019sentencebert,gao2021simcse}. A shared relation can already affect that ranking. For example, the theater history may score above unrelated histories with similarly low lexical scores, while a set of pantry-related distractors still keeps it below the retrieval cutoff. This is a consequence of letting all candidates compete in one ranking. The underlying distinction between shared relations and surface resemblance is familiar from research on analogy \cite{gentner1983structure,forbus1995macfac}.

In \emph{passive associative retrieval}, a description of the current situation serves as the query. It requires neither an explicit retrieval question nor the name of the desired relation. We use lexical scores to organize the candidates and retain the frozen encoder's scores for ranking. In the pantry example, hard conditioning gives the theater history priority over the paperwork history while keeping its dense-score advantage over unrelated histories with similarly low lexical scores. Soft conditioning uses a finite score correction based on lexical similarity.

We analyze when a history below the global cutoff can enter the top ranks and when subtracting a baseline can reverse a useful ordering. In the controlled pools, the number of candidates with low lexical scores exceeds the number of positions. Hard priority alone therefore leaves a selection problem within that group. Score permutations test whether the gain depends on which dense score belongs to each history. Competitor replacements test the ranking against histories with similar but distinct interventions. Input probes examine how the problem description affects dense-score margins.

The study shows how the current situation can bring relevant experiences into a limited reading budget by changing which histories compete. A shared library also reveals the cost of hard priority: related histories with similar wording can lose their places. The observed trade-off motivates keeping an ordinary resemblance-based route alongside the associative one.

\section{Related Work}\label{sec:related}
\subsection{Relational similarity and memory access}
The distinction between shared relations and shared surface content originates in work on analogy. Structure-mapping theory emphasizes systematic relational correspondence \cite{gentner1983structure}; MAC/FAC separates inexpensive content-based access from more detailed structural comparison \cite{forbus1995macfac}. Contemporary evaluations retain this distinction: ARN varies surface and system mappings \cite{sourati2024arn}, while AnaloBench examines abstract and long-context analogies \cite{ye2024analobench}. These lines of work motivate our retrieval objective. We investigate an earlier access decision: whether a relation-bearing history enters a finite candidate budget when its frozen representation already provides some useful discrimination.

\subsection{Representations, rank fusion, and conditional scores}
Lexical retrieval and learned representations offer different measures of similarity. BM25 models term matching \cite{robertson2009bm25}, whereas dense passage and sentence representations support semantic matching \cite{karpukhin2020dpr,reimers2019sentencebert,gao2021simcse}. Comparisons of sparse, dense, and attentional representations \cite{luan2021sparse}, together with BEIR and MTEB \cite{thakur2021beir,muennighoff2023mteb}, establish the importance of the retrieval setting in assessing these representations. Combining ranked lists through reciprocal rank fusion \cite{cormack2009rrf} provides an alternative to choosing one score. Our intervention acts on competition itself: lexical conditions determine which candidates compete first, and dense scores discriminate within those conditions.

Subtracting a reference score also has precedents. Cross-domain similarity local scaling corrects neighborhood-density effects in cross-lingual matching \cite{conneau2018csls}. The residual formulation studied here instead conditions on an explicit lexical covariate. Its conditional mean interpretation explains what the corrected score measures, but does not guarantee relation-preserving rankings. We examine that distinction through candidate margins and finite-budget retrieval.

\subsection{Associative access in agent memory}
Agent memory involves both maintenance and access. Long-term conversational memory studies information updates \cite{bae2022memory}; Generative Agents combines stored experiences, reflection, and dynamic retrieval \cite{park2023agents}. HippoRAG and its successor organize associative access through graph structure and personalized PageRank \cite{gutierrez2024hippo,gutierrez2025hippo2}, while AssoMem uses clue-based memory graphs and multiple retrieval signals \cite{zhang2025assomem}. These approaches show that associative memory is already an established research direction. Surface conditioning contributes a particular mechanism for organizing competition among externally stored histories with fixed representations.

LongMemEval distinguishes indexing, retrieval, and reading \cite{wu2025longmem}; our measurements concern retrieval before downstream reading. Context-gated associative retrieval in Hopfield and transformer memory studies contextual effects within the memory dynamics \cite{choraria2026gate}. Here the context supplies scores for external histories, and lexical conditioning changes their ranking. These settings motivate related questions about access while requiring different mechanisms and evidence.

\section{Surface-Conditioned Associative Retrieval}\label{sec:method}

\subsection{Retrieval objective and observable conditions}\label{sec:problem}

Let $q$ be a current situation and $\mathcal H=\{x_i\}_{i=1}^{N}$ a history library. The context contains state, observations, constraints, and recent actions; it need not include a retrieval question directed at a known target. A history describes a problem, the intervention taken, its outcome, and the resulting lesson. An association label $y^A(q,x)$ indicates a transferable relation or intervention pathway. Surface resemblance concerns entities, topics, and wording. Association and resemblance may coincide or separate.

\begin{definition}[Passive associative retrieval]
Given a current situation $q$ without an explicit retrieval question and a history library $\mathcal H$, return a candidate set $\mathcal C(q)$ within budget $k$. The objective is to increase access to histories with transferable value, especially those differing in surface content, while supporting ordinary resemblance-based access where needed.
\end{definition}
Exposure and adoption are different decisions. Retrieval lets a downstream process inspect whether the past premises and intervention apply to the present constraints. An association label does not authorize copying an action without that check. The experiments evaluate candidate access and ranking; long-term agent use motivates the objective.

\begin{table}[tbp]
\centering\small
\caption{Surface resemblance and associative value. The low-surface stratum must contain related and unrelated competitors.}
\label{tab:quadrants}
\begin{tabularx}{\linewidth}{p{2.6cm}XX}
\toprule
 & Related & Unrelated\\
\midrule
High surface & Similar topic and shared mechanism & Similar entities or wording, different intervention\\
Low surface & Different topic, transferable relation & Different surface content without the target relation\\
\bottomrule
\end{tabularx}
\end{table}
Table~\ref{tab:quadrants} separates two attributes of a history: resemblance to the current situation and the availability of a transferable relation. Lexical scores determine the retrieval strata, while relation labels are used to evaluate the returned histories. A history reused under a different context is assessed against that context's relation.

Let $L(q,x)\in[0,1]$ be a lexical score and $S(q,x)=\cos(e(q),e(x))\in[-1,1]$ the cosine similarity of frozen embeddings. The lexical covariate may use term overlap or another explicit surface estimator. We use BM25 followed by empirical-CDF normalization against calibration scores. This is an observable proxy, not a complete decomposition of semantics.

A separate background set $\mathcal B$ provides calibration pairs $(L_b,S_b)$. Its median lexical score defines $\tau$, and its scores fit $f$. The background determines the lexical reference, boundary, and baseline; study contexts do not enter their fitting. Embedding weights remain fixed throughout retrieval. The construction and selection of study materials are described in Section~\ref{sec:experiments}.

\subsection{When lexical priority changes the top ranks}\label{sec:hard}\label{sec:mechanism}

Define the low-lexical-score stratum $\mathcal H_\ell(q)=\{x:L(q,x)\leq\tau\}$. Hard-stratum ranking first selects by $S$ from this stratum, then fills any remaining positions by $S$ from the other stratum. When the prioritized stratum exceeds the budget, related and unrelated histories still compete on their dense scores.

For dense scores bounded by $[-1,1]$, choose $\lambda>2$. Ranking by
\begin{equation}
A_{\mathrm{hard}}(q,x)=S(q,x)-\lambda\ind[L(q,x)>\tau]
\end{equation}
prioritizes the low-lexical stratum and preserves dense order within each stratum. For any same-stratum pair $p,n$, the score difference remains $S(q,p)-S(q,n)$.

Every low-stratum score is at least $-1$, whereas every high-stratum score is at most $1-\lambda<-1$. Within a stratum the penalty is constant and cancels in a difference.

The implementation uses the equivalent score $S+4\ind[L\leq\tau]$. The constant 4 enforces priority across the entire cosine range rather than tuning a fusion weight. The calibrated boundary is 0.5 in this study. Priority preserves available within-stratum distinctions; useful retrieval still requires those distinctions to correspond to association.

\begin{algorithm}[H]
\caption{Hard-stratum retrieval under a fixed budget}\label{alg:primary}
\begin{algorithmic}[1]
\Require Context $q$, histories $\mathcal H$, fixed scores $L,S$, boundary $\tau$, budget $k$
\State $\mathcal H_\ell\gets\{x\in\mathcal H:L(q,x)\leq\tau\}$
\State $\mathcal C\gets\Top_{\min(k,|\mathcal H_\ell|)}(\mathcal H_\ell,S)$
\If{$|\mathcal C|<\min(k,|\mathcal H|)$}
\State Fill from $\mathcal H\setminus\mathcal H_\ell$ by $S$ up to $\min(k,|\mathcal H|)$ items
\EndIf
\State \Return $\mathcal C$, retaining scores, strata, and ranks
\end{algorithmic}
\end{algorithm}

Suppose two histories have low lexical scores and the related one $p$ has higher dense similarity than the unrelated one $n$. Higher-scoring surface distractors $h$ can still fill the global top $k$. Hard conditioning lets $p$ and $n$ compete before $h$ while retaining their dense-score order. Ranking solely by increasing lexical score can place $n$ first.

\begin{proposition}[Budget exposure]\label{prop:release}
Fix a common tie order. For $p\in\mathcal H_\ell$, let $r_S(p)$ be its global dense rank and $r_\ell(p)$ its dense rank within the low-lexical stratum. Hard-stratum ranking selects $p$ if and only if $r_\ell(p)\leq\min(k,|\mathcal H_\ell|)$. When $|\mathcal H_\ell|\geq k$,
\begin{equation}\label{eq:release}
r_S(p)>k,\qquad r_\ell(p)\leq k
\end{equation}
is necessary and sufficient for a history excluded by dense Top-$k$ to be exposed by hard conditioning.
\end{proposition}
\begin{proof}
Hard prioritization places all low-stratum candidates first in their dense order. The selected low-stratum items are exactly its first $\min(k,|\mathcal H_\ell|)$ candidates. Combining this with dense exclusion, $r_S(p)>k$, gives the result.
\end{proof}
The target reaches the top $k$ only if it ranks high enough within the prioritized stratum. With fewer than $k$ candidates in that stratum, all of them enter automatically. The competition pool in Section~\ref{sec:experiments} puts more candidates in the prioritized stratum than there are positions, so the dense-score order affects which histories are retrieved.

A minimal structure contains a low-lexical related item $p$, low-lexical unrelated item $n$, and high-lexical unrelated item $h$. If $S_h>S_p>S_n$ and $L_n<L_p\leq\tau<L_h$, dense ranking selects $h$, inverse lexical ranking selects $n$, and hard conditioning selects $p$. For a baseline that is constant in each stratum, residual ranking also selects $p$ when the difference between the two constants exceeds $S_h-S_p$. Section~\ref{sec:cases} connects these inequalities to histories with identifiable intervention pathways. The full-pool experiments evaluate their retrieval under competition beyond a three-history example.

\FloatBarrier
\subsection{Conditional baselines and ranking margins}

The soft formulation is
\begin{equation}\label{eq:residual}
A(q,x)=S(q,x)-f(L(q,x)),\qquad f(L)\approx\mathbb E[S\mid L].
\end{equation}
The baseline can use ordinary least squares (OLS), low-capacity binned interpolation, or stratum constants. Subtracting the mean within each of two strata preserves within-stratum order but imposes only a finite cross-stratum penalty. This estimates a background baseline; hard ranking instead specifies an access priority.

If $f(L)=\mathbb E[S\mid L]$ and the expectations exist, then $\mathbb E[A\mid L]=0$ by conditional-expectation linearity. A positive residual measures similarity above the conditional background mean.
Centering does not guarantee optimal association ranking. The population conditional mean, its finite-sample estimate, and the task's desired ranking are different objects. A reasonable background estimate can remove distinctions that are useful for the target relation; candidate-level score differences reveal when this occurs.

For related $p$ and unrelated $n$, define $\Delta S=S_p-S_n$ and $\Delta f=f(L_p)-f(L_n)$. Then
\begin{equation}\label{eq:decomposition}
\Delta A=\Delta S-\Delta f.
\end{equation}
\begin{proposition}[Rank-reversal margins]\label{prop:flip}
Under strict score comparison, a positive dense margin $\Delta S>0$ is reversed exactly when $\Delta f>\Delta S$. A negative margin $\Delta S<0$ becomes positive exactly when $\Delta f<\Delta S$. Equality produces a tie resolved by the fixed tie order.
\end{proposition}
\begin{proof}
Substitute $\Delta A<0$ or $\Delta A>0$ into Equation~\ref{eq:decomposition} and rearrange.
\end{proof}
Useful subtraction must be strong enough across strata and restrained within them. Stratum-constant subtraction has $\Delta f=0$ within a stratum, preserving useful dense margins, but its cross-stratum mean difference may be insufficient to overcome distractors. Finer bins allow local correction while exposing useful pairs to unequal penalties. If that difference exceeds their dense margin, ranking deteriorates. Calibration capacity therefore matters through these margins, not simply through the number of bins.

\subsection{Complementary access and computational cost}

The framework can retain a resemblance channel $\mathcal C_R=\Top_{k_R}(\mathcal H,L)$ and an associative channel $\mathcal C_A=\Top_{k_A}(\mathcal H,A)$, using either hard or soft $A$. Deduplication retains source channels and scores. Any evaluation must match the final unique-candidate and downstream judgment budgets and record duplicate handling and refill rules. The present results motivate this extension through complementary selections; they do not evaluate a final channel-allocation policy.

\begin{figure}[tbp]
\centering
\begin{tikzpicture}[>=Latex,font=\small,
 flow/.style={->,semithick,draw=blue!27!black,rounded corners=3pt},
 extension/.style={flow,dashed,draw=black!58},
 card/.style={rounded corners=3pt,inner xsep=5pt,inner ysep=4pt,align=center,draw=blue!20!black!45,text=blue!24!black},
 score/.style={card,minimum width=3.6cm},
 heading/.style={font=\small\bfseries,text=blue!28!black}]
\node[card,fill=blue!8,minimum width=9.75cm] (query) at (5.03,3.34)
  {\textbf{Current situation $q$} \enspace+\enspace \textbf{history library $\mathcal H$}};
\node[card,fill=black!3,minimum width=4.65cm] (background) at (12.78,2.03)
  {\textbf{Separate calibration set}\\{\footnotesize Cutoff $\tau$ and baseline $f(L)$}};
\node[score,fill=orange!12] (lexical) at (2.44,2.03)
  {\textbf{Lexical score $L$}\\{\footnotesize BM25 with a fixed reference}};
\node[score,fill=teal!10] (dense) at (7.83,2.03)
  {\textbf{Frozen dense score $S$}\\{\footnotesize Embedding cosine similarity}};
\draw[flow] (2.44,3.03) -- (lexical.north);
\draw[flow] (7.83,3.03) -- (dense.north);

\draw[rounded corners=5pt,fill=blue!4,draw=blue!18!black!27] (.04,-.99) rectangle (4.84,1.06);
\draw[rounded corners=5pt,fill=blue!4,draw=blue!18!black!27] (5.12,-.99) rectangle (15.22,1.06);
\node[heading] at (2.44,.74) {Resemblance ranking};
\node[text=blue!24!black,align=center] at (2.44,-.13)
  {Rank by $L$\\Higher lexical scores first};
\node[heading] at (10.18,.74) {Associative ranking};
\node[card,fill=orange!10,text width=4.15cm] at (7.72,-.21)
  {\textbf{Hard priority}\\{\footnotesize Low-$L$ group first}\\{\footnotesize Rank each group by $S$}};
\node[card,fill=teal!10,text width=4.15cm] at (12.66,-.21)
  {\textbf{Soft correction}\\{\footnotesize Subtract the lexical baseline}\\{\footnotesize Rank by $S-f(L)$}};
\draw[flow] (lexical.south) -- (2.44,1.06);
\draw[flow] (lexical.east) -- (5.43,2.03) -- (5.43,1.06);
\draw[flow] (dense.south) -- (7.83,1.06);
\draw[extension] (background.south) -- (12.78,1.06);

\draw[rounded corners=5pt,fill=teal!6,draw=teal!30!black!39] (.04,-2.45) rectangle (15.22,-1.51);
\node[heading] at (7.63,-1.78) {Proposed two-channel retrieval};
\node[font=\footnotesize,text=blue!24!black] at (7.63,-2.16)
  {Merge and deduplicate \quad $\cdot$ \quad Keep source labels and scores \quad $\cdot$ \quad Share one reading budget};
\draw[extension] (2.44,-.99) -- (2.44,-1.51);
\draw[extension] (10.18,-.99) -- (10.18,-1.51);
\end{tikzpicture}
\caption{Context-triggered retrieval from a history library. Resemblance ranking uses lexical scores. Associative ranking uses the same lexical scores to set priorities or adjust frozen dense scores. The experiments evaluate the rankings separately. Their complementary coverage motivates the proposed combination under one reading budget.}
\label{fig:framework}
\end{figure}
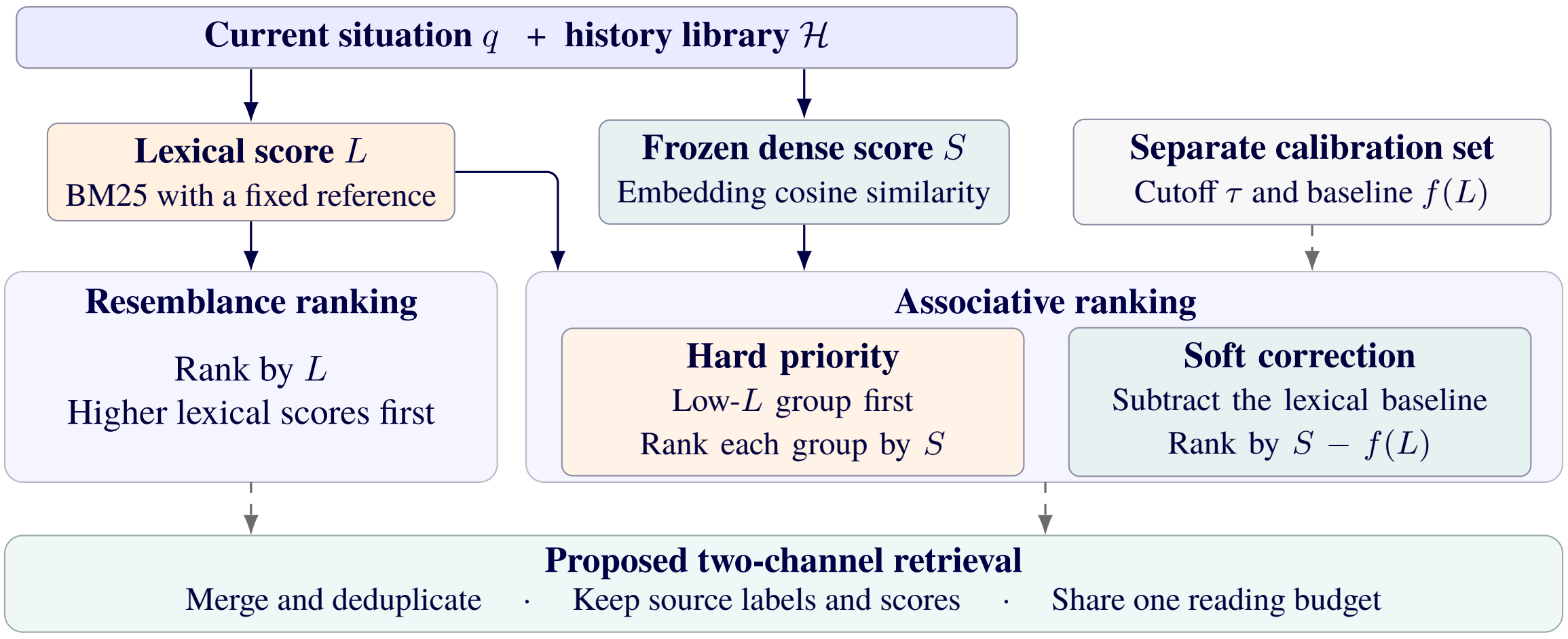
Figure~\ref{fig:framework} places the access mechanism before downstream use. With cached vectors and a lexical index, exact scoring computes $L$ and $S$ over $N$ candidates; a heap selects Top-$k$ in $O(N\log k)$ time after scoring. Conditioning adds a stratum test or scalar baseline evaluation, without per-candidate generation calls. A large-library implementation can condition a preliminary candidate union, but must evaluate any recall lost before reranking. Here all candidate pools are scored exactly.

\section{Experimental Design}\label{sec:experiments}
The principal competition pool puts 18 candidates with low lexical scores in contention for ten positions per context. Some share the intervention with the current situation, and the others do not. This design tests whether dense similarity can select the relevant histories once they receive lexical priority.

\subsection{Data and experimental units}
The exploratory dataset contains eight contexts and 144 candidate records: four calibration contexts with 48 histories and four study contexts with 96. Calibration relations concern admission control and positional re-anchoring. Study relations concern arrival staggering and completion-evidence recovery, each instantiated in everyday planning and software troubleshooting. The four study contexts instantiate two relation specifications, not four independent relations.

We study two relations. \emph{Arrival staggering} disperses synchronized demand while holding service capacity fixed, as in a single-counter pantry, synchronized database refreshes, or theater coat collection. \emph{Completion-evidence recovery} restores verified evidence of work already completed, as in baking checklists, software dependency records, or nursery shipping approvals. Adding service capacity differs from staggering; performing an inspection for the first time differs from restoring evidence of a completed inspection.

Texts were primarily generated and reviewed using Doubao Seed 2.1 Turbo. Review concealed candidate labels while providing the relation specification; it used the same model and was not independent annotation. Later cross-query labels were checked by reading intervention pathways and prerequisites. Retry backoff with jitter, for example, may support staggering but not completion-evidence recovery. Six histories with ambiguous staggering relevance are excluded from derived staggering pools; evidence-recovery pools retain their own relation checks.

The retrieval counts use query--history pairs. A history can appear under multiple queries, so pairs can share a history and an intervention pattern. Scores informed material construction and the selection of illustrative cases. The comparisons are descriptive and use the two constructed intervention patterns. Appendix~\ref{app:exploration} documents the design changes and selection sequence.

\subsection{Fixed representations and lexical references}
We use BGE-base-en-v1.5 \cite{bge} with CLS pooling, normalized float32 vectors, and cosine similarity. Histories include descriptions of the problem, action, outcome, and lesson; contexts use the retrieval prefix. Inputs are checked against the 512-token limit before encoding, avoiding silent truncation. Exact model versions and encoding settings are documented in the accompanying reproducibility materials.

The three-pool comparison uses all 48 calibration histories as a common BM25 corpus, with its empirical CDF and fitted baselines fixed for the study contexts. Two additional reference conditions isolate the effect of changing corpus statistics: one recomputes BM25 as distractors are added, and the other fixes each query's initial 12-item corpus and retains its original calibrator. Those stress analyses use their own surface labels and are reported separately from the common-reference comparison.

\subsection{Candidate pools, comparisons, and metrics}
The \emph{original pool} retains each context's histories, with fixed-reference regrouping and relation-specific ambiguity exclusions. The \emph{competition pool} combines six distant related histories from the same relation across domains, 12 low-surface unrelated backgrounds, and the context's surface distractors. Its low-lexical stratum has 18 candidates per context. The \emph{shared library} starts from all 96 study histories, applying query-specific relation checks and ambiguity exclusions.
\begin{table}[htbp]
\centering\footnotesize
\caption{Candidate-pool composition. Per-context sizes follow the order staggering/everyday, staggering/software, evidence/everyday, evidence/software. Surface strata use the fixed calibration reference; all counts are descriptive.}\label{tab:pools}
\begin{tabular}{llrrrrr}
\toprule
Pool & Sizes by context & Occurrences & Low/rel. & High/rel. & Low/unrel. & High/unrel.\\
\midrule
Original pool & 24/23/24/24 & 95 & 18 & 6 & 12 & 59\\
Competition pool & 33/32/33/33 & 131 & 24 & 0 & 48 & 59\\
Shared library & 90/90/96/96 & 372 & 44 & 6 & 263 & 59\\
\bottomrule
\end{tabular}
\end{table}

\FloatBarrier
Every method selects 10 candidates per context using the same available $L,S$ and fixed tie order. References include BM25, dense similarity, RRF with constant 60 \cite{cormack2009rrf}, inverse raw BM25, OLS residuals, 4-bin and 20-bin interpolated residuals, and a two-stratum residual. Boundaries and fitters use the original 48 calibration pairs, without parameter search on study contexts. Unrelated-only calibration is a separate label-informed diagnostic. Inverse raw BM25 and inverse lexical-CDF ranking differ when normalization creates additional ties; they are identified separately in the relevant experiments.

For related sets $P_q$, associative recall and precision are
\begin{equation}
R_A=\frac{\sum_q|\mathcal C_q\cap P_q|}{\sum_q|P_q|},\qquad
P_A=\frac{\sum_q|\mathcal C_q\cap P_q|}{\sum_q|\mathcal C_q|}.
\end{equation}
Low- and high-surface related recall use their actual available targets; an absent category is a dash rather than zero. Unrelated selections are split by surface stratum.

\section{Results}\label{sec:results}

\subsection{Retrieval under a fixed selection budget}

Table~\ref{tab:main} compares access to related histories at the same selection budget. The competition pool directly tests the proposed mechanism because its distant candidates outnumber the available positions; the original pools and shared library examine how the result changes with the candidates that compete.
\begin{table}[htbp]
\centering\small
\caption{Associative retrieval across three pools. Each cell gives retrieved/relevant query--history occurrences, with precision in parentheses. Every method selects 40 occurrences per pool. Column winners, including ties and reference methods, are bold; bold does not indicate significance.}\label{tab:main}
\begin{tabular}{lrrr}
\toprule
Method & Original pool & Competition pool & Shared library\\
\midrule
BM25 & 5/24 (12.5\%) & 0/24 (0.0\%) & 5/50 (12.5\%)\\
Dense & 3/24 (7.5\%) & 0/24 (0.0\%) & 3/50 (7.5\%)\\
Hybrid RRF & 3/24 (7.5\%) & 0/24 (0.0\%) & 3/50 (7.5\%)\\
Inverse BM25 & 19/24 (47.5\%) & 4/24 (10.0\%) & 2/50 (5.0\%)\\
OLS residual & 7/24 (17.5\%) & 1/24 (2.5\%) & 7/50 (17.5\%)\\
4-bin residual & 13/24 (32.5\%) & 4/24 (10.0\%) & 8/50 (20.0\%)\\
20-bin residual & 13/24 (32.5\%) & 8/24 (20.0\%) & 10/50 (25.0\%)\\
Two-stratum residual & 15/24 (37.5\%) & 14/24 (35.0\%) & \best{13/50 (32.5\%)}\\
\textbf{Hard-stratum ranking} & \best{21/24 (52.5\%)}& \best{21/24 (52.5\%)}& \best{13/50 (32.5\%)}\\
\bottomrule
\end{tabular}
\end{table}

In the competition pool, hard-stratum ranking retrieves 21 of 24 relevant query--history pairs, giving recall of 87.5\% and precision of 52.5\% (21 of 40 selections). Dense ranking and RRF retrieve none, and inverse BM25 retrieves four. Giving low-lexical-score candidates priority brings them into contention. Ranking them by dense similarity then retrieves more relevant pairs than inverse BM25. Two-stratum subtraction retrieves 14, 20-bin subtraction eight, and OLS one. A constant correction can preserve the order within the low-lexical-score stratum, as the two-stratum residual does, yet leave the cross-stratum penalty too small to displace candidates with higher lexical scores.

\subsubsection{Interpreting the retrieved relations}\label{sec:cases}

Two examples show what the retrieved relation contributes. They were selected after inspecting scores to illustrate the mechanism, as detailed in Appendix~\ref{app:exploration}. Figure~\ref{fig:ranks} summarizes their ranking changes; Appendix~\ref{app:case_scores} reports the underlying scores. Complete source histories are preserved in the accompanying case materials.

\paragraph{Arrival staggering.}
The pantry and theater differ in setting but share synchronized arrivals at a single service point. Staggered coat-collection windows change arrival timing while retaining the lone attendant, matching the pantry's capacity constraint. The donation-record history is lexically closer to the pantry, yet its intervention restores access to paperwork and does not redistribute arrivals. A distant violin-repair history addresses a different physical cause. The distinction that matters for transfer is therefore the intervention on demand timing under fixed capacity.

\paragraph{Completion-evidence recovery.}
The baking problem involves completed preparation whose missing checklist marks block later work. In the related nursery history, 12 trays have already passed inspection, but absent grading cards prevent dispatch. Recovering verified approvals restores evidence of completed work and permits release without another inspection. The surface-similar oven history changes electrical capacity by splitting the workload, while violin repair changes a physical component. Neither addresses the missing evidence. Here the shared premise that the prerequisite has already been completed determines which intervention transfers.

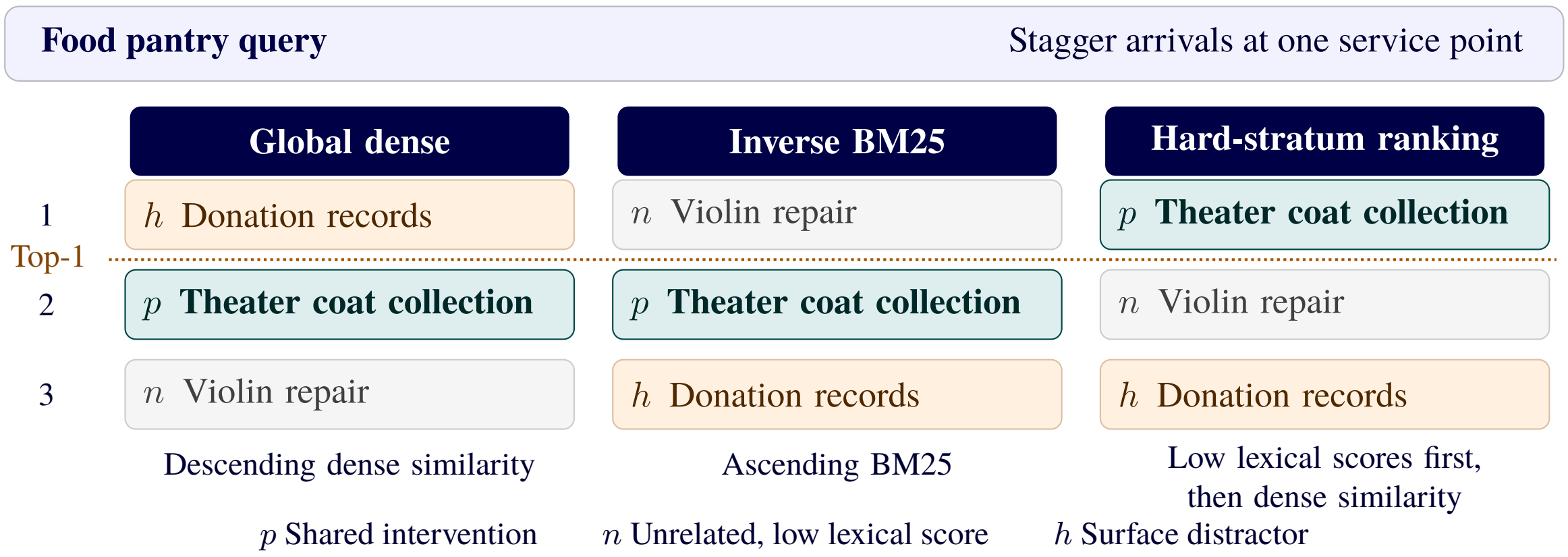
\begin{figure}[htbp]
\centering
\begin{tikzpicture}[font=\small,
 method/.style={rounded corners=3pt,minimum height=6.8mm,text width=4.05cm,align=center,fill=blue!28!black,text=white,font=\small\bfseries},
 entry/.style={rounded corners=3pt,minimum height=6.8mm,text width=4.05cm,align=left,inner xsep=5pt,draw=black!20},
 p/.style={fill=teal!13,draw=teal!55!black,text=teal!31!black,font=\small\bfseries},
 n/.style={fill=black!4,text=black!75},
 h/.style={fill=orange!12,draw=orange!58!black!35,text=orange!31!black},
 key/.style={font=\footnotesize,text=blue!20!black},
 reason/.style={key,text width=4.05cm,align=center}]
\draw[rounded corners=4pt,fill=blue!5,draw=blue!16!black!28] (0,2.43) rectangle (15.28,3.16);
\node[anchor=west,text=blue!28!black,font=\small\bfseries] at (.22,2.80) {Food pantry query};
\node[anchor=east,text=blue!20!black] at (15.02,2.80) {Stagger arrivals at one service point};
\node[method] at (3.38,1.84) {Global dense};
\node[method] at (8.16,1.84) {Inverse BM25};
\node[method] at (12.94,1.84) {Hard-stratum ranking};
\foreach \rank in {1,2,3} {
  \pgfmathsetmacro{\row}{1.12-.88*(\rank-1)}
  \node[key] at (.41,\row) {\rank};
}
\foreach \role/\history/\denseRank/\inverseRank/\hardRank in {
  p/{Theater coat collection}/2/2/1,
  n/{Violin repair}/3/1/2,
  h/{Donation records}/1/3/3
} {
  \foreach \column/\rank in {3.38/\denseRank,8.16/\inverseRank,12.94/\hardRank} {
    \pgfmathsetmacro{\row}{1.12-.88*(\rank-1)}
    \node[entry,\role] at (\column,\row) {$\role$\enspace \history};
  }
}
\draw[densely dotted,thick,draw=orange!64!black] (1.02,.68) -- (15.21,.68);
\node[key,text=orange!52!black] at (.43,.68) {Top-1};
\node[reason] at (3.38,-1.35) {Descending dense similarity};
\node[reason] at (8.16,-1.35) {Ascending BM25};
\node[reason] at (12.94,-1.48) {Low lexical scores first,\\then dense similarity};
\node[key] at (7.64,-2.03) {$p$ Shared intervention\qquad $n$ Unrelated, low lexical score\qquad $h$ Surface distractor};
\end{tikzpicture}
\caption{Rank changes in the three-candidate food-pantry example. The line marks a one-position budget within the triple. The theater history shares the intervention and moves from rank two under global dense ranking to rank one under hard conditioning. The evidence-recovery triple has the same ordering by candidate role. Table~\ref{tab:main} reports the separate full-pool results at ten selections per context.}\label{fig:ranks}
\end{figure}

In both examples, the related history $p$ outranks the unrelated distant history $n$ by dense similarity, while the surface distractor $h$ outranks both. Dense order is $h>p>n$, inverse lexical order is $n>p>h$, and hard-conditioned order is $p>n>h$. Thus, the representation supplies a useful distinction that global selection fails to expose. The two-stratum residual also places $p$ first because its cross-stratum penalty exceeds the distractor's advantage while leaving the $p$--$n$ margin unchanged. These score-selected examples explain the access pathway; the full competition results in Table~\ref{tab:main} establish its behavior with more distant candidates than available positions.

\subsection{Dense discrimination and conditional calibration}\label{sec:attribution}

\subsubsection{Candidate-specific dense information}

If the gain came entirely from stratum membership, it would persist when candidate-specific dense information was removed. With uniform selection within the prioritized stratum, the expected count is 13.33 related occurrences, below the observed 21. This expectation avoids information carried by deterministic record ordering. Inverse raw BM25 tests a different lexical preference and retrieves four.

Permuting the correspondence between candidates and their dense scores retains the candidate set, threshold, and score multiset. Across 20 fixed seeds, broad-stratum permutations average 14.1 related selections (range 10--19); permutations within at most four blocks of adjacent unique lexical levels average 17.2 (13--19); permutations within exactly equal $L$ groups average 19.65 (19--21). Equal lexical levels are never split between blocks.
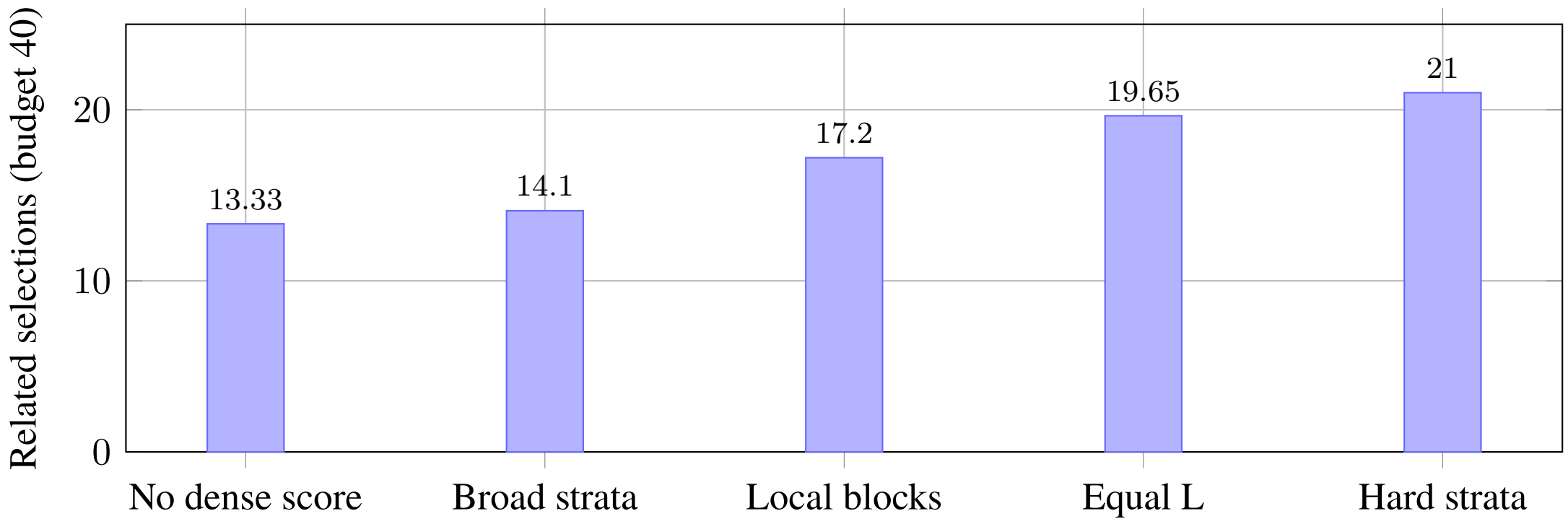
\begin{figure}[htbp]
\centering
\begin{tikzpicture}\begin{axis}[width=.91\linewidth,height=5.5cm,ymin=0,ymax=25,xtick={1,2,3,4,5},xticklabels={No dense score,Broad strata,Local blocks,Equal L,Hard strata},ylabel={Related selections (budget 40)},ybar,bar width=20pt,nodes near coords,every node near coord/.append style={font=\scriptsize},grid=major]
\addplot[fill=blue!30,draw=blue!60] coordinates {(1.000000,13.333333) (2.000000,14.100000) (3.000000,17.200000) (4.000000,19.650000) (5.000000,21.000000)};
\end{axis}\end{tikzpicture}
\caption{Information ablations of hard-stratum retrieval. Permutation bars average 20 fixed seeds. The no-dense bar is the uniform tie-selection expectation, avoiding quadrant-encoded candidate identifiers.}\label{fig:ablation}
\end{figure}

Broad-stratum permutation preserves lexical priority and the dense-score distribution, but reduces retrieval relative to the original 21 selections (Figure~\ref{fig:ablation}). The correspondence between a history and its dense score therefore contributes to the gain. Narrower permutations retain more lexical and label composition, approaching the original count. Only one or two equal-$L$ groups per context contain both related and unrelated histories, limiting how much those permutations can change. The comparisons identify information retained by different exchange rules; their seeds describe variation on the same material and do not constitute independent relation tests.

\subsubsection{Conditional calibration and preserved margins}

In the originally labeled low-surface subset, each context has three related and three unrelated histories. Dense Top-3 retrieves 9/12 related occurrences; OLS residuals retrieve 4/12, 4-bin residuals 6/12, and 20-bin residuals 8/12. The representation already contains useful coarse discrimination, while detailed subtraction can disturb its order. Hard conditioning preserves that order; stratum constants test a lower-capacity finite penalty.

The calibration diagnostics examine how stable the subtracted baseline is when its supporting material changes. Leaving out one calibration context while holding the original CDF fixed changes all-background OLS predictions by at most about 0.027 on a 101-point grid, 4-bin predictions by 0.045, and 20-bin predictions by 0.176. Under the later global reference, low/high stratum means are 0.5708/0.8771, each supported by 24 calibration examples; their maximum leave-context-out mean change is 0.0121. These describe estimation behavior rather than a direct measure of retrieval quality.

Changing calibration to unrelated-only background does not consistently improve retrieval. At maximum stress with fixed original-pool statistics, OLS total related selections remain 11, 4-bin selections change from 12 to seven, and 20-bin selections from nine to 10. Because background selection reads association labels, this analysis diagnoses calibration effects rather than providing an implemented label-free background selector. Appendix~\ref{app:calibration} retains curves, occupancies, and sensitivity values.

\subsubsection{Score references under changing competition}

Adding distractors can change both ranks and BM25 corpus statistics. Under a dynamic reference, even unchanged texts and dense scores can receive different $L$ values and penalties. At maximum stress, fixing the original-pool statistics changes 20-bin low-surface retrieval from 2/12 to 6/12 and OLS from 0/12 to 3/12. Fixing the reference removes that source of drift, without resolving every competition effect.
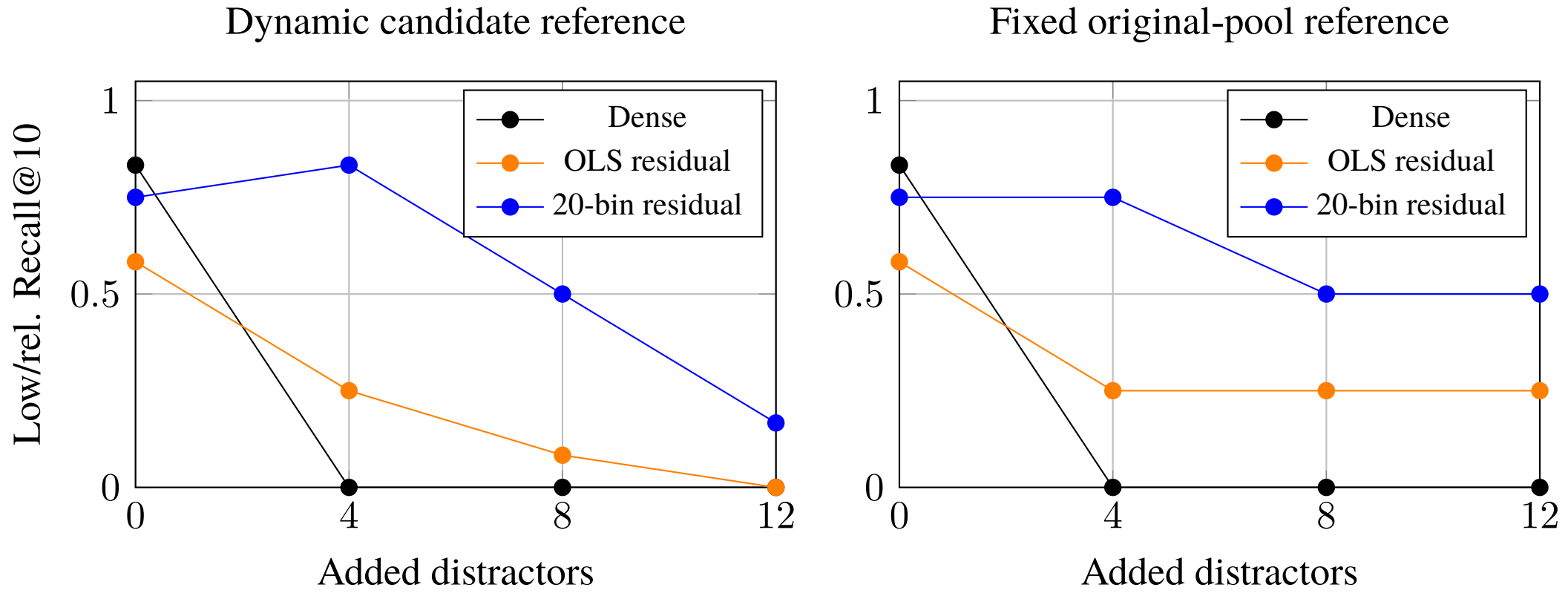
\begin{figure}[htbp]
\centering
\begin{tikzpicture}\begin{groupplot}[group style={group size=2 by 1,horizontal sep=1.1cm},width=.45\linewidth,height=5.2cm,ymin=0,ymax=1.05,xmin=0,xmax=12,xtick={0,4,8,12},grid=major,xlabel={Added distractors},ylabel={Low/rel. Recall@10},legend style={font=\scriptsize}]
\nextgroupplot[title={Dynamic candidate reference}]
\addplot[black,mark=*] coordinates {(0.000000,0.833333) (4.000000,0.000000) (8.000000,0.000000) (12.000000,0.000000)};\addlegendentry{Dense}
\addplot[orange,mark=*] coordinates {(0.000000,0.583333) (4.000000,0.250000) (8.000000,0.083333) (12.000000,0.000000)};\addlegendentry{OLS residual}
\addplot[blue,mark=*] coordinates {(0.000000,0.750000) (4.000000,0.833333) (8.000000,0.500000) (12.000000,0.166667)};\addlegendentry{20-bin residual}
\nextgroupplot[title={Fixed original-pool reference},ylabel={}]
\addplot[black,mark=*] coordinates {(0.000000,0.833333) (4.000000,0.000000) (8.000000,0.000000) (12.000000,0.000000)};\addlegendentry{Dense}
\addplot[orange,mark=*] coordinates {(0.000000,0.583333) (4.000000,0.250000) (8.000000,0.250000) (12.000000,0.250000)};\addlegendentry{OLS residual}
\addplot[blue,mark=*] coordinates {(0.000000,0.750000) (4.000000,0.750000) (8.000000,0.500000) (12.000000,0.500000)};\addlegendentry{20-bin residual}
\end{groupplot}\end{tikzpicture}
\caption{Separating score-reference drift from competition in the original stress study. The fixed reference uses each query's original 12 candidates and retains the original calibrator. The later three-pool experiment uses a separate global 48-history reference.}\label{fig:pressure}
\end{figure}

Figure~\ref{fig:pressure} uses the original stress-study surface labels. Table~\ref{tab:main} uses strata recomputed under global calibration. The former isolates reference drift; the latter evaluates the constructed retrieval pools.

\FloatBarrier
\subsection{Prerequisites and competition-induced errors}

Paired mechanism probes compare an original related history, a same-mechanism paraphrase, a changed mechanism, and its paraphrase for each relation. The staggering counterexample adds parallel service capacity instead of changing arrival timing. The evidence-recovery counterexample performs previously uncompleted inspections instead of restoring existing completion evidence. Counterexamples share the related history's topic and much wording; they need not share the current query's topic. The two phrasings belong to one counterexample family.

Using a fixed identifier order, we replace one and then two original low-surface unrelated backgrounds per context. Removal does not use model rank. Candidate count, relevant count, stratum sizes, representation, calibration, and Top-10 budget remain fixed; retained candidates keep their scores. Hard-stratum retrieval changes from 21 to 21 to 19, while dense, BM25, and RRF remain at zero and the two-stratum residual remains at 14. Appendix~\ref{app:metrics} gives every method.

Both new misses concern evidence recovery: each context changes from 5/6 to 4/6. The staggering contexts remain at 5/6 and 6/6. The concentration of new misses in evidence recovery identifies a difficult distinction: completing a prerequisite and restoring evidence that it is already complete. Most of the original retrieval gain survives this stronger competition. The probes had already been inspected during exploration. The replacement analysis therefore measures a specified change in competition using known material.

\subsubsection{A related history displaced by stronger competitors}

The original competition result includes 19 unrelated selections alongside 21 related ones. They include a lapsed-insurance history, retrieved in all four contexts, in which payment restores genuinely expired coverage. This differs from recovering evidence that a prerequisite already holds. Other false selections concern physical repair. Thus, original low-surface backgrounds already include coarse recovery semantics that can exceed fine mechanism distinctions; they are not all easy negatives.

The first new miss in query/candidate-ID order concerns a CI release. Upstream tests succeeded, but a registry outage lost their attestations; verified signed artifacts can restore the release evidence. Its dense score remains 0.6100, while its rank changes from 10 to 11 after two replacements. Newly added first-time inspection histories score about 0.667 and 0.670 and enter the budget.
\begin{table}[htbp]
\centering\small
\caption{A budget-boundary change in the evidence/everyday context. The original related history retains its scores but moves from rank 10 to 11 after the second mechanism-altered replacement.}\label{tab:failure}
\begin{tabular}{lrrrrr}
\toprule
History & L & S & Original rank & Replace 1 & Replace 2\\
\midrule
Related history & 0.4792 & 0.6100 & 10 & 10 & 11\\
Perform inspection & 0.3333 & 0.6667 & --- & 4 & 5\\
Inspection paraphrase & 0.1458 & 0.6702 & --- & --- & 4\\
\bottomrule
\end{tabular}
\end{table}

\FloatBarrier
The target's score has not deteriorated. Stronger within-stratum competition removes its position. Conditioning supplies access opportunities, while precise premise discrimination still depends on the representation and downstream checks. Existing gains and identified errors jointly inform the method: preserve prerequisites, identify the intervention object, and distinguish completion from evidence recovery.

\subsubsection{The role of prerequisite information}

The two relations depend on different information in a history. Arrival staggering is expressed directly in the action, whereas evidence recovery also requires the premise that work is already complete. To examine this distinction, histories are re-encoded using the action and lesson or the problem and action, with query representations fixed. Action-and-lesson inputs increase staggering margins but make evidence-recovery margins negative. Problem-and-action inputs restore positive original-versus-changed margins in all four contexts (Table~\ref{tab:input}).
\begin{table}[htbp]
\centering\small
\caption{Dense-score margins between an original related history and its mechanism-altered counterpart. Only history input changes; query representations remain fixed. Positive values favor the related history. The largest margin in each context is bold. These probes diagnose representations, not full retrieval performance.}\label{tab:input}
\begin{tabular}{lrrr}
\toprule
Context & Full text & Action+Lesson & Problem+Action\\
\midrule
Staggering/everyday & +0.0097 & \best{+0.0215} & +0.0142\\
Staggering/software & +0.0124 & \best{+0.0214} & +0.0134\\
Evidence/everyday & \best{+0.0100} & -0.0494 & +0.0077\\
Evidence/software & +0.0016 & -0.0080 & \best{+0.0166}\\
\bottomrule
\end{tabular}
\end{table}

For evidence recovery, the problem description can show that an inspection already took place. An action description may omit that fact, leaving the encoder less able to distinguish a first inspection from recovery of its records. Including the problem description restores a positive margin in these probes. The input changes content, length, and wording together. The observed effect is on embedding-similarity margins, and we did not rerun full-pool retrieval with these input variants.

\FloatBarrier
\subsection{Which histories each candidate pool favors}\label{sec:coverage}
In the original pool, hard conditioning retrieves 21/24, dense ranking 3/24, and inverse BM25 19/24. In the shared library, the denominator increases to 50: hard conditioning and two-stratum subtraction each retrieve 13, dense ranking three, and inverse BM25 two. The change from 21 to 13 is not a decline on a common denominator; available related histories, background composition, and cross-relation competition all change.
\begin{table}[htbp]
\centering\small
\caption{Composition of hard-stratum retrieval. Relevant columns give retrieved/available occurrences; unrelated columns give selected counts. These are composition summaries across different pools, not a common-denominator comparison.}\label{tab:composition}
\begin{tabular}{lrrrr}
\toprule
Pool & Low/rel. & High/rel. & Low/unrel. & High/unrel.\\
\midrule
Original pool & 18/18 & 3/6 & 12 & 7\\
Competition pool & 21/24 & --- & 19 & 0\\
Shared library & 13/44 & 0/6 & 27 & 0\\
\bottomrule
\end{tabular}
\end{table}

In the original pools, hard conditioning retrieves all 18 relevant pairs with low lexical scores and three of the six with high scores. In the shared library, the prioritized stratum fills every position, leaving six relevant pairs with high lexical scores outside the top ranks. All relevant pairs in the competition pool have low lexical scores, so that pool tests a different aspect of the ranking.

\section{Discussion}\label{sec:discussion}
\subsection{What conditioning preserves in the dense ranking}
Hard conditioning determines which histories compete for the first positions. Their dense scores determine which ones win those positions. Shuffling scores across histories within the same lexical stratum reduces the number of relevant selections, confirming that the lexical partition accounts for only part of the result. The input probes identify a separate difficulty for the encoder: evidence-recovery histories are easier to distinguish from first-time inspections when the history includes the problem description.

Subtracting a baseline places each dense score relative to a lexical reference. Its effect on the ranking depends on how much the corrections differ across candidates. A finer fit can reverse the order of two histories whose dense scores were already useful for the task. A constant correction within each stratum preserves that order, though the difference between the two constants may be too small to move a relevant history into the top ranks.

\subsection{Complementary access and downstream use}
In the shared library, BM25 retrieves five of the six relevant pairs with high lexical scores and none of the 44 with low lexical scores. Hard conditioning retrieves 13 of the low-lexical-score pairs and none of the six with high scores. This complementary coverage motivates the two channels in Figure~\ref{fig:framework}. Evaluating their combination requires a shared budget of unique histories and rules for handling duplicates and filling open positions.

For an agent, a change in task state or an attempted recovery could trigger access using a summary of the current situation. Returned histories would then be examined for matching prerequisites, intervention objects, and capacity constraints. The distinction between doing missing work and recovering evidence of completed work illustrates why retrieval must be followed by this judgment. Candidate exposure is the outcome measured here; improvements in reasoning and action remain to be evaluated.

\subsection{Scope and further research}
The evidence comes from a constructive study of two relation specifications in four study contexts with one frozen encoder. Scores informed material construction and illustrative-case selection. The same model generated and reviewed most texts, reused histories produce correlated query--history occurrences, and the mechanism probes were inspected before the replacement analysis. These conditions support an account of the realized mechanism, with descriptive comparisons rather than population estimates or independent held-out validation of relations.

Natural histories introduce variation in relevance prevalence, overlapping relations, background composition, and the expression of prerequisites. Further study can fix the access method before evaluating new material, vary lexical estimators and encoders, and connect candidate exposure to downstream decisions. The present analyses identify useful questions for that extension: whether the representation retains the relevant distinction, whether conditioning preserves it, and whether stronger within-stratum competition prevents it from reaching the available budget.

\section{Conclusion}
An experience can rank above unrelated histories with similarly low lexical scores and still miss the global retrieval cutoff. Hard conditioning lets the current situation bring that experience into contention while preserving the order supplied by the frozen encoder. Soft subtraction offers a graded correction whose effects can be traced to pairwise score margins. The shared-library results motivate pairing the associative ranking with a resemblance-based route, so histories that share wording and histories that share an intervention can both be considered for subsequent reasoning.
\FloatBarrier

\appendix
\section{Calibration and Score Diagnostics}\label{app:calibration}
\subsection{Fitted baselines and sample support}
The calibration plots show the background observations, OLS fit, 4/20-bin interpolation, and two-stratum means under the global 48-history reference. Bin occupancy indicates the sample support available to each calibration function.
\begin{figure}[htbp]
\centering
\begin{tikzpicture}\begin{axis}[width=.88\linewidth,height=6cm,xmin=0,xmax=1,ymin=.4,ymax=1,xlabel={$L$},ylabel={$f(L)$},title={All background},legend style={at={(.02,.98)},anchor=north west,font=\scriptsize},grid=major]
\addplot[orange,thick] coordinates {(0.000000,0.478303) (0.025000,0.490588) (0.050000,0.502872) (0.075000,0.515156) (0.100000,0.527440) (0.125000,0.539724) (0.150000,0.552008) (0.175000,0.564292) (0.200000,0.576577) (0.225000,0.588861) (0.250000,0.601145) (0.275000,0.613429) (0.300000,0.625713) (0.325000,0.637997) (0.350000,0.650282) (0.375000,0.662566) (0.400000,0.674850) (0.425000,0.687134) (0.450000,0.699418) (0.475000,0.711702) (0.500000,0.723986) (0.525000,0.736271) (0.550000,0.748555) (0.575000,0.760839) (0.600000,0.773123) (0.625000,0.785407) (0.650000,0.797691) (0.675000,0.809976) (0.700000,0.822260) (0.725000,0.834544) (0.750000,0.846828) (0.775000,0.859112) (0.800000,0.871396) (0.825000,0.883680) (0.850000,0.895965) (0.875000,0.908249) (0.900000,0.920533) (0.925000,0.932817) (0.950000,0.945101) (0.975000,0.957385) (1.000000,0.969670)};\addlegendentry{OLS}
\addplot[blue,thick] coordinates {(0.000000,0.552933) (0.025000,0.552933) (0.050000,0.552933) (0.075000,0.552933) (0.100000,0.552933) (0.125000,0.552933) (0.150000,0.556515) (0.175000,0.560097) (0.200000,0.563679) (0.225000,0.567261) (0.250000,0.570843) (0.275000,0.574425) (0.300000,0.578007) (0.325000,0.581589) (0.350000,0.585171) (0.375000,0.588753) (0.400000,0.615373) (0.425000,0.641992) (0.450000,0.668612) (0.475000,0.695231) (0.500000,0.721850) (0.525000,0.748470) (0.550000,0.775089) (0.575000,0.801709) (0.600000,0.828328) (0.625000,0.854947) (0.650000,0.859384) (0.675000,0.863820) (0.700000,0.868257) (0.725000,0.872693) (0.750000,0.877129) (0.775000,0.881566) (0.800000,0.886002) (0.825000,0.890439) (0.850000,0.894875) (0.875000,0.899312) (0.900000,0.899312) (0.925000,0.899312) (0.950000,0.899312) (0.975000,0.899312) (1.000000,0.899312)};\addlegendentry{4-bin interpolation}
\addplot[red,thick] coordinates {(0.000000,0.586399) (0.025000,0.586399) (0.050000,0.593172) (0.075000,0.602201) (0.100000,0.572111) (0.125000,0.522462) (0.150000,0.518590) (0.175000,0.547415) (0.200000,0.548770) (0.225000,0.522653) (0.250000,0.552561) (0.275000,0.593673) (0.300000,0.602335) (0.325000,0.600180) (0.350000,0.579914) (0.375000,0.550592) (0.400000,0.575540) (0.425000,0.600487) (0.450000,0.589421) (0.475000,0.575080) (0.500000,0.709284) (0.525000,0.845132) (0.550000,0.840662) (0.575000,0.837188) (0.600000,0.844666) (0.625000,0.852144) (0.650000,0.860756) (0.675000,0.867625) (0.700000,0.871008) (0.725000,0.872268) (0.750000,0.867158) (0.775000,0.865287) (0.800000,0.879612) (0.825000,0.885444) (0.850000,0.882783) (0.875000,0.887868) (0.900000,0.903800) (0.925000,0.918506) (0.950000,0.930761) (0.975000,0.939952) (1.000000,0.939952)};\addlegendentry{20-bin interpolation}
\addplot[black,dashed,thick] coordinates {(0.000000,0.570843) (0.500000,0.570843) (0.500001,0.877129) (1.000000,0.877129)};\addlegendentry{Two-stratum constant}
\addplot[only marks,mark=*,mark size=1pt,gray] coordinates {(0.927083,0.909713) (0.656250,0.860621) (0.593750,0.811084) (0.406250,0.606636) (0.468750,0.549355) (0.489583,0.596025) (0.697917,0.841307) (0.864583,0.863842) (0.822917,0.841031) (0.010417,0.565560) (0.385417,0.593447) (0.156250,0.547146) (0.760417,0.904805) (0.947917,0.932315) (0.989583,0.928309) (0.197917,0.582030) (0.427083,0.580553) (0.239583,0.521703) (0.718750,0.881090) (0.552083,0.872165) (0.635417,0.861091) (0.135417,0.432825) (0.447917,0.620510) (0.302083,0.583982) (0.885417,0.891597) (0.843750,0.888252) (0.739583,0.898242) (0.343750,0.590808) (0.322917,0.608116) (0.281250,0.643484) (0.781250,0.820994) (0.531250,0.829159) (0.572917,0.826445) (0.031250,0.591419) (0.072917,0.607247) (0.052083,0.602219) (0.968750,0.959233) (0.510417,0.862595) (0.677083,0.872373) (0.260417,0.584388) (0.093750,0.603175) (0.218750,0.514898) (0.614583,0.843198) (0.802083,0.932520) (0.906250,0.919129) (0.364583,0.507737) (0.177083,0.541625) (0.114583,0.525353)};\addlegendentry{48 calibration pairs}
\end{axis}\end{tikzpicture}
\caption{Calibration curves using all 48 background pairs under the global fixed reference. Curves show the fitted calibration functions.}\label{fig:calall}
\end{figure}
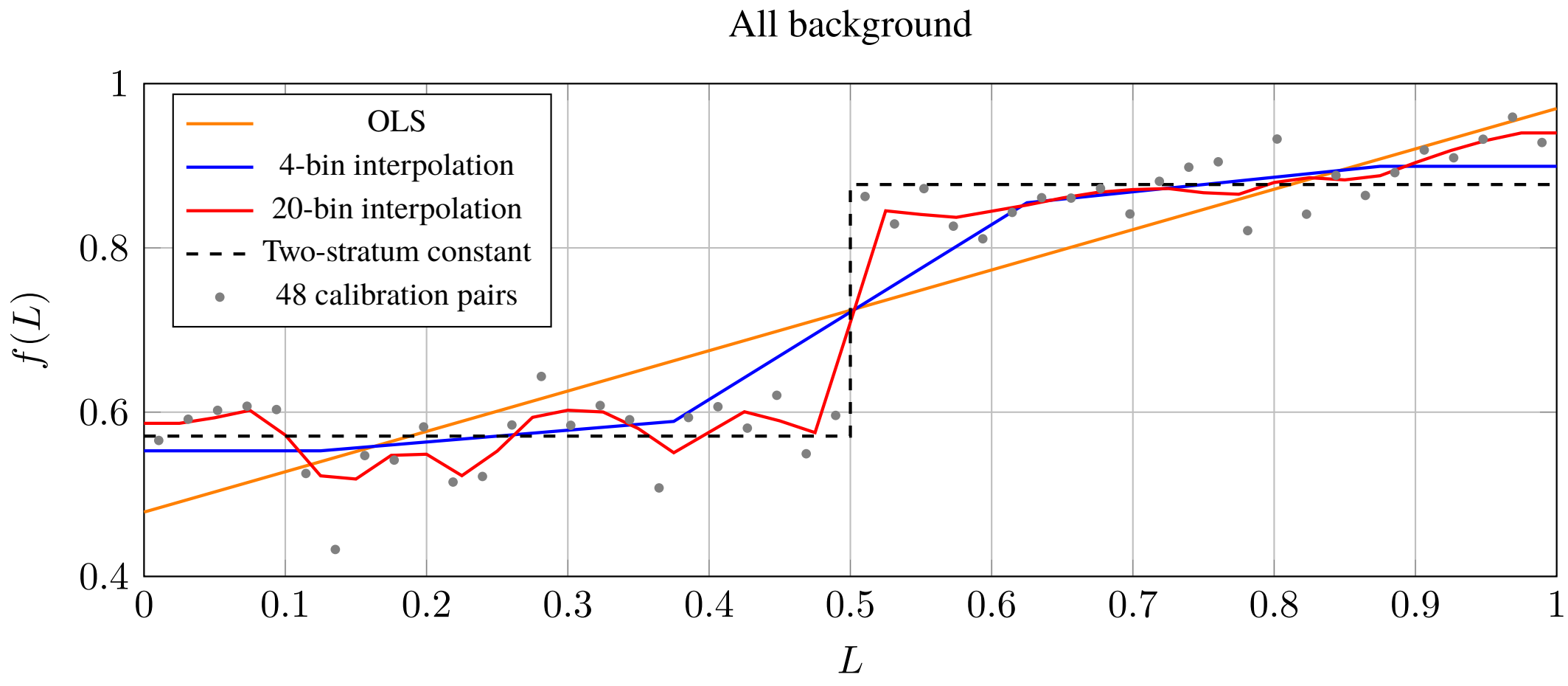
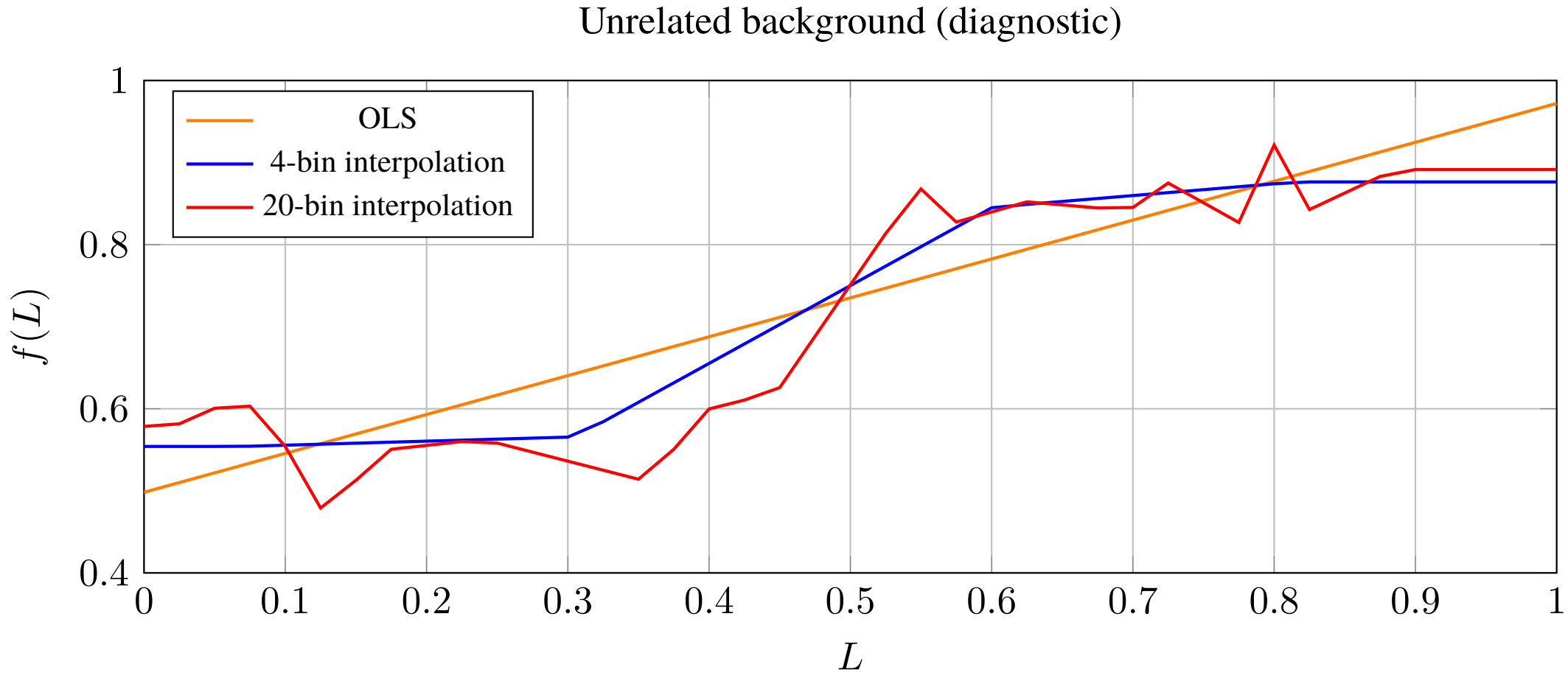
\begin{figure}[htbp]
\centering
\begin{tikzpicture}\begin{axis}[width=.88\linewidth,height=6cm,xmin=0,xmax=1,ymin=.4,ymax=1,xlabel={$L$},ylabel={$f(L)$},title={Unrelated background (diagnostic)},legend style={at={(.02,.98)},anchor=north west,font=\scriptsize},grid=major]
\addplot[orange,thick] coordinates {(0.000000,0.498229) (0.025000,0.510075) (0.050000,0.521920) (0.075000,0.533766) (0.100000,0.545611) (0.125000,0.557457) (0.150000,0.569302) (0.175000,0.581147) (0.200000,0.592993) (0.225000,0.604838) (0.250000,0.616684) (0.275000,0.628529) (0.300000,0.640375) (0.325000,0.652220) (0.350000,0.664065) (0.375000,0.675911) (0.400000,0.687756) (0.425000,0.699602) (0.450000,0.711447) (0.475000,0.723293) (0.500000,0.735138) (0.525000,0.746983) (0.550000,0.758829) (0.575000,0.770674) (0.600000,0.782520) (0.625000,0.794365) (0.650000,0.806211) (0.675000,0.818056) (0.700000,0.829901) (0.725000,0.841747) (0.750000,0.853592) (0.775000,0.865438) (0.800000,0.877283) (0.825000,0.889129) (0.850000,0.900974) (0.875000,0.912820) (0.900000,0.924665) (0.925000,0.936510) (0.950000,0.948356) (0.975000,0.960201) (1.000000,0.972047)};\addlegendentry{OLS}
\addplot[blue,thick] coordinates {(0.000000,0.554104) (0.025000,0.554104) (0.050000,0.554104) (0.075000,0.554378) (0.100000,0.555610) (0.125000,0.556842) (0.150000,0.558074) (0.175000,0.559307) (0.200000,0.560539) (0.225000,0.561771) (0.250000,0.563003) (0.275000,0.564235) (0.300000,0.565467) (0.325000,0.584176) (0.350000,0.607879) (0.375000,0.631581) (0.400000,0.655283) (0.425000,0.678986) (0.450000,0.702688) (0.475000,0.726391) (0.500000,0.750093) (0.525000,0.773795) (0.550000,0.797498) (0.575000,0.821200) (0.600000,0.844902) (0.625000,0.849046) (0.650000,0.852632) (0.675000,0.856217) (0.700000,0.859802) (0.725000,0.863388) (0.750000,0.866973) (0.775000,0.870558) (0.800000,0.874144) (0.825000,0.876434) (0.850000,0.876434) (0.875000,0.876434) (0.900000,0.876434) (0.925000,0.876434) (0.950000,0.876434) (0.975000,0.876434) (1.000000,0.876434)};\addlegendentry{4-bin interpolation}
\addplot[red,thick] coordinates {(0.000000,0.578489) (0.025000,0.581653) (0.050000,0.600637) (0.075000,0.603152) (0.100000,0.554016) (0.125000,0.479089) (0.150000,0.512850) (0.175000,0.550669) (0.200000,0.555366) (0.225000,0.560064) (0.250000,0.558215) (0.275000,0.547201) (0.300000,0.536188) (0.325000,0.525175) (0.350000,0.514161) (0.375000,0.550592) (0.400000,0.599762) (0.425000,0.610587) (0.450000,0.625726) (0.475000,0.688321) (0.500000,0.750916) (0.525000,0.813511) (0.550000,0.867864) (0.575000,0.827473) (0.600000,0.839808) (0.625000,0.852144) (0.650000,0.848429) (0.675000,0.844713) (0.700000,0.845285) (0.725000,0.875080) (0.750000,0.851042) (0.775000,0.827004) (0.800000,0.921367) (0.825000,0.842712) (0.850000,0.862894) (0.875000,0.883076) (0.900000,0.891486) (0.925000,0.891486) (0.950000,0.891486) (0.975000,0.891486) (1.000000,0.891486)};\addlegendentry{20-bin interpolation}
\end{axis}\end{tikzpicture}
\caption{Calibration curves using unrelated background pairs. This is a label-informed diagnostic under the global reference.}\label{fig:calnegative}
\end{figure}

\begin{table}[htbp]
\centering\footnotesize
\caption{Occupancy of the 20 calibration bins. Intervals are left-closed and right-open, with 1 included in the last bin. Predictions interpolate between stored nonempty-bin centers. Counts and means are descriptive.}\label{tab:bins}
\begin{tabular}{rrrr}
\toprule
Bin & L interval & Count & Mean S\\
\midrule
1 & 0.00--0.05 & 2 & 0.5785\\
2 & 0.05--0.10 & 3 & 0.6042\\
3 & 0.10--0.15 & 2 & 0.4791\\
4 & 0.15--0.20 & 3 & 0.5569\\
5 & 0.20--0.25 & 2 & 0.5183\\
6 & 0.25--0.30 & 2 & 0.6139\\
7 & 0.30--0.35 & 3 & 0.5943\\
8 & 0.35--0.40 & 2 & 0.5506\\
9 & 0.40--0.45 & 3 & 0.6026\\
10 & 0.45--0.50 & 2 & 0.5727\\
11 & 0.50--0.55 & 2 & 0.8459\\
12 & 0.55--0.60 & 3 & 0.8366\\
13 & 0.60--0.65 & 2 & 0.8521\\
14 & 0.65--0.70 & 3 & 0.8581\\
15 & 0.70--0.75 & 2 & 0.8897\\
16 & 0.75--0.80 & 2 & 0.8629\\
17 & 0.80--0.85 & 3 & 0.8873\\
18 & 0.85--0.90 & 2 & 0.8777\\
19 & 0.90--0.95 & 3 & 0.9204\\
20 & 0.95--1.00 & 2 & 0.9438\\
\bottomrule
\end{tabular}
\end{table}

\FloatBarrier
\subsection{Sensitivity to the calibration context}
Removing one calibration context measures the stability of the fitted baseline. Table~\ref{tab:sensitivity} holds the original empirical CDF fixed and evaluates maximum change on a 101-point grid. Table~\ref{tab:loo} gives the analogous stratum means under the global reference. These are sensitivity analyses of their respective estimators, not held-out retrieval evaluations.
\begin{table}[htbp]
\centering\small
\caption{Original-calibrator sensitivity to leaving out one calibration context. The original empirical CDF is fixed; maximum prediction change is evaluated on a 101-point grid. This differs from the later global-reference calibration.}\label{tab:sensitivity}
\begin{tabular}{lrr}
\toprule
Fitter & All background & Unrelated background\\
\midrule
OLS & 0.0269 & 0.0370\\
4 bins & 0.0455 & 0.0576\\
20 bins & 0.1758 & 0.0944\\
\bottomrule
\end{tabular}
\end{table}

\begin{table}[htbp]
\centering\small
\caption{Two-stratum calibration after leaving out one background context. The empirical CDF and boundary are fixed. Sensitivity describes estimation behavior, not retrieval quality.}\label{tab:loo}
\begin{tabular}{lrrr}
\toprule
Omitted context & Low mean & High mean & Max. change\\
\midrule
Admission/everyday & 0.5690 & 0.8846 & 0.0075\\
Admission/software & 0.5766 & 0.8706 & 0.0065\\
Re-anchoring/everyday & 0.5587 & 0.8831 & 0.0121\\
Re-anchoring/software & 0.5791 & 0.8701 & 0.0082\\
\bottomrule
\end{tabular}
\end{table}

\FloatBarrier
\subsection{Within-stratum ranking and background selection}
Table~\ref{tab:top3} holds the candidate subset fixed to examine how subtraction changes useful dense order. Table~\ref{tab:background} varies the calibration background under the two original stress references. Selecting unrelated backgrounds uses relation labels and serves as a diagnostic of the baseline, not as a label-free retrieval procedure.
\begin{table}[htbp]
\centering\small
\caption{Original low-surface subset diagnostic: six items per context, three related and three unrelated, with Top-3 selection. The largest retrieved count is bold. Original surface labels differ from later CDF-based strata.}\label{tab:top3}
\begin{tabular}{lr}
\toprule
Ranking & Related selected\\
\midrule
Dense & \best{9/12}\\
Inverse lexical CDF & 4/12\\
OLS residual & 4/12\\
4-bin residual & 6/12\\
20-bin residual & 8/12\\
\bottomrule
\end{tabular}
\end{table}

\begin{table}[htbp]
\centering\footnotesize
\caption{Calibration-background comparison at maximum stress. Unrelated-only calibration uses relationship labels and is diagnostic. References are shown separately; comparisons pair all-background and unrelated-background versions of the same fitter. Larger paired values, including ties, are bold.}\label{tab:background}
\begin{tabular}{llrrrr}
\toprule
Reference & Fitter & All: low recall & Unrel.: low recall & All: related & Unrel.: related\\
\midrule
Dynamic & OLS & \best{0/12} & \best{0/12} & \best{1} & \best{1}\\
Dynamic & 4 bins & 1/12 & \best{2/12} & \best{2} & \best{2}\\
Dynamic & 20 bins & 2/12 & \best{3/12} & \best{6} & 4\\
Fixed original & OLS & \best{3/12} & \best{3/12} & \best{11} & \best{11}\\
Fixed original & 4 bins & \best{5/12} & 3/12 & \best{12} & 7\\
Fixed original & 20 bins & 6/12 & \best{7/12} & 9 & \best{10}\\
\bottomrule
\end{tabular}
\end{table}

\FloatBarrier
\section{Retrieval Composition and Competition}\label{app:metrics}
\subsection{Coverage within each candidate pool}
Tables~\ref{tab:full1}--\ref{tab:full3} separate related recall from unrelated selections in each lexical stratum. They explain the complementary coverage discussed in Section~\ref{sec:coverage}; candidate composition and relevant denominators differ across pools.
\begin{table}[htbp]
\centering\small
\caption{Original-pool retrieval composition. Related columns are higher-is-better; unrelated counts are lower-is-better. Column winners and ties are bold. A dash means no related item of that type is available.}\label{tab:full1}
\begin{tabular}{lrrrr}
\toprule
Method & Low/rel. & High/rel. & Low/unrel. & High/unrel.\\
\midrule
BM25 & 0/18 & \best{5/6}& \best{0}& 35\\
Dense & 0/18 & 3/6 & \best{0}& 37\\
Hybrid RRF & 0/18 & 3/6 & \best{0}& 37\\
Inverse BM25 & \best{18/18}& 1/6 & 12 & 9\\
OLS residual & 6/18 & 1/6 & 2 & 31\\
4-bin residual & 9/18 & 4/6 & 4 & 23\\
20-bin residual & 11/18 & 2/6 & 5 & 22\\
Two-stratum residual & 12/18 & 3/6 & 5 & 20\\
\textbf{Hard-stratum ranking} & \best{18/18}& 3/6 & 12 & \best{7}\\
\bottomrule
\end{tabular}
\end{table}
\begin{table}[htbp]
\centering\small
\caption{Competition-pool retrieval composition. Related columns are higher-is-better; unrelated counts are lower-is-better. Column winners and ties are bold. A dash means the type is absent.}\label{tab:full2}
\begin{tabular}{lrrrr}
\toprule
Method & Low/rel. & High/rel. & Low/unrel. & High/unrel.\\
\midrule
BM25 & 0/24 & --- & \best{0}& 40\\
Dense & 0/24 & --- & \best{0}& 40\\
Hybrid RRF & 0/24 & --- & \best{0}& 40\\
Inverse BM25 & 4/24 & --- & 36 & \best{0}\\
OLS residual & 1/24 & --- & 6 & 33\\
4-bin residual & 4/24 & --- & 7 & 29\\
20-bin residual & 8/24 & --- & 9 & 23\\
Two-stratum residual & 14/24 & --- & 10 & 16\\
\textbf{Hard-stratum ranking} & \best{21/24}& --- & 19 & \best{0}\\
\bottomrule
\end{tabular}
\end{table}
\begin{table}[htbp]
\centering\small
\caption{Shared-library retrieval composition. Related columns are higher-is-better; unrelated counts are lower-is-better. Column winners and ties are bold.}\label{tab:full3}
\begin{tabular}{lrrrr}
\toprule
Method & Low/rel. & High/rel. & Low/unrel. & High/unrel.\\
\midrule
BM25 & 0/44 & \best{5/6}& \best{0}& 35\\
Dense & 0/44 & 3/6 & \best{0}& 37\\
Hybrid RRF & 0/44 & 3/6 & \best{0}& 37\\
Inverse BM25 & 2/44 & 0/6 & 38 & \best{0}\\
OLS residual & 6/44 & 1/6 & 9 & 24\\
4-bin residual & 8/44 & 0/6 & 27 & 5\\
20-bin residual & 10/44 & 0/6 & 30 & \best{0}\\
Two-stratum residual & \best{13/44}& 0/6 & 27 & \best{0}\\
\textbf{Hard-stratum ranking} & \best{13/44}& 0/6 & 27 & \best{0}\\
\bottomrule
\end{tabular}
\end{table}

\FloatBarrier
\subsection{Mechanism-altered competition}
Table~\ref{tab:contexts} locates the replacement effects by context, and Table~\ref{tab:replacement} compares methods under the same relevant denominator and selection budget. This separates the survival of the overall gain from the particular contexts in which new competitors displace a target.
\begin{table}[htbp]
\centering\small
\caption{Hard-stratum retrieval by context in the competition pool. Related counts are out of six per context; the final column gives unrelated selections in the original pool. Counts across replacement versions are observations, not model selection scores.}\label{tab:contexts}
\begin{tabular}{lrrrrr}
\toprule
Context & Pool size & Original & Replace 1 & Replace 2 & Unrelated\\
\midrule
Staggering/everyday & 33 & 5/6 & 5/6 & 5/6 & 5\\
Staggering/software & 32 & 6/6 & 6/6 & 6/6 & 4\\
Evidence/everyday & 33 & 5/6 & 5/6 & 4/6 & 5\\
Evidence/software & 33 & 5/6 & 5/6 & 4/6 & 5\\
\bottomrule
\end{tabular}
\end{table}

\begin{table}[htbp]
\centering\small
\caption{Related occurrences retrieved after replacing one or two unrelated backgrounds per context. The relevant denominator remains 24 and the total selection budget remains 40. Column winners are bold.}\label{tab:replacement}
\begin{tabular}{lrrr}
\toprule
Method & Original & Replace 1 & Replace 2\\
\midrule
BM25 & 0 & 0 & 0\\
Dense & 0 & 0 & 0\\
Hybrid RRF & 0 & 0 & 0\\
Inverse BM25 & 4 & 4 & 4\\
OLS residual & 1 & 1 & 1\\
4-bin residual & 4 & 4 & 4\\
20-bin residual & 8 & 8 & 8\\
Two-stratum residual & 14 & 14 & 14\\
\textbf{Hard-stratum ranking} & \best{21}& \best{21}& \best{19}\\
\bottomrule
\end{tabular}
\end{table}

\FloatBarrier
\noindent\begin{minipage}{\linewidth}
\subsection{Constraints on permutation comparisons}
Table~\ref{tab:groups} describes how many histories can exchange scores within exact lexical ties. Few mixed-label groups restrict the perturbations available to the equal-$L$ diagnostic, which is why its near-original retrieval count must be interpreted together with the broader permutations.
\begin{table}[H]
\centering\small
\caption{Available mixing within equal-lexical-score permutation groups. Every context has 18 low-lexical candidates. Counts describe the restricted exchange space.}\label{tab:groups}
\begin{tabular}{lrrr}
\toprule
Context & Unique L groups & Mixed groups & Items in mixed groups\\
\midrule
Staggering/everyday & 11 & 2 & 5\\
Staggering/software & 9 & 2 & 10\\
Evidence/everyday & 11 & 1 & 4\\
Evidence/software & 12 & 1 & 3\\
\bottomrule
\end{tabular}
\end{table}

\end{minipage}

\FloatBarrier
\section{Exploratory Design and Interpretation}\label{app:exploration}
The study developed through score-informed construction and diagnostic analysis. Initial stress experiments varied the candidate pool while recomputing lexical statistics. Comparing dynamic and fixed references identified both score-reference drift and competition effects; changing the calibration background did not improve every fitter. These findings motivated the separate global-reference comparisons reported in the main results. They are successive analyses of constructed material, not independent replications.

Illustrative triples were selected after examining dense and lexical margins. Screening found 687 eligible combinations for staggering and 740 for evidence recovery; these are construction inventories rather than independent trials. A subsequent constraint required the first-ranked distractor to exceed the calibration median. Clearer triples resulted, but both versions retained identical full-pool Top-10 scores. The main case analysis uses the first qualifying construction per relation in query-identifier order. Complete texts and selection mappings are preserved in the accompanying research materials.

Changing history inputs clarified the role of prerequisites. Action+Lesson inputs increased staggering margins but reversed evidence-recovery margins. Problem+Action inputs restored positive original-versus-changed margins in all four contexts. These probes altered content, length, and wording together. They informed the mechanism analysis and were not evaluated as new full-method retrieval systems with the old full-text calibrator.

Under the global reference and full history inputs, two-stratum subtraction selected 14 related occurrences in the competition pool and hard conditioning selected 21. Replacing one and then two unrelated backgrounds by mechanism-altered histories produced hard-conditioning counts of 21 and 19. The comparisons retain the same representations and scores for unchanged histories. Because the changed-mechanism probes had already informed exploration, replacement measures a specified change in competition rather than performance on an untouched test set.

An early local-block permutation split equal lexical scores across groups, yielding mean 18 and range 14--22. The corrected analysis groups unique lexical levels without splitting ties and adds permutations within exactly equal lexical scores. This changes which histories may exchange scores in the diagnostic, while leaving the retrieval scores and the 21/21/19 replacement sequence unchanged. The complete exploration sequence, including five length-related and six surface-related diagnostic flags, is preserved in the accompanying materials.

\FloatBarrier
\section{Scores for the Illustrative Cases}\label{app:case_scores}
The two score-selected cases in Section~\ref{sec:cases} share the same ranking pattern but instantiate different transferable relations. Tables~\ref{tab:caseT} and~\ref{tab:caseE} provide the lexical score $L$, dense score $S$, two-stratum baseline $f(L)$, and residual. In each case the related and unrelated distant histories receive the same baseline, preserving their dense margin. The surface distractor receives a larger baseline, sufficient to move it below the related history. These tables support the mechanism interpretation in Figure~\ref{fig:ranks}; they are not additional retrieval trials.

\begin{table}[htbp]
\centering\small
\caption{Three-candidate staggering example. Hard-stratum ranking and the two-stratum residual both select the related history. Coordinates are descriptive; larger scores do not imply better histories across columns.}\label{tab:caseT}
\begin{tabular}{llrrrr}
\toprule
History & Role & L & S & f(L) & S-f(L)\\
\midrule
Coat collection & Related target & 0.5000 & 0.6270 & 0.5708 & +0.0562\\
Violin repair & Low-surface unrelated & 0.3542 & 0.4275 & 0.5708 & -0.1433\\
Donation records & Surface distractor & 0.6042 & 0.7882 & 0.8771 & -0.0889\\
\bottomrule
\end{tabular}
\end{table}

\begin{table}[htbp]
\centering\small
\caption{Three-candidate evidence-recovery example. Hard-stratum ranking and the two-stratum residual both select the related history.}\label{tab:caseE}
\begin{tabular}{llrrrr}
\toprule
History & Role & L & S & f(L) & S-f(L)\\
\midrule
Nursery dispatch & Related target & 0.4792 & 0.6929 & 0.5708 & +0.1221\\
Violin repair & Low-surface unrelated & 0.3542 & 0.5363 & 0.5708 & -0.0345\\
Oven capacity & Surface distractor & 0.6042 & 0.8619 & 0.8771 & -0.0152\\
\bottomrule
\end{tabular}
\end{table}

\FloatBarrier
\clearpage
\begingroup
\small
\raggedright

\endgroup
\end{document}